# Punctuated evolution of influenza virus hemagglutinin (A/H1N1) under opposing migration and vaccination pressures

J. C. Phillips

Dept. of Physics and Astronomy, Rutgers University, Piscataway, N. J., 08854

1-908-273-8218     jcphillips8@comcast.net

Abstract

Influenza virus contains two highly variable envelope glycoproteins, hemagglutinin (HA) and neuraminidase (NA).  The structure and properties of HA, which is responsible for binding the virus to the cell that is being infected, change significantly when the virus is transmitted from avian or swine species to humans.  Previously we identified much smaller human individual evolutionary amino acid mutational changes in NA, which cleaves sialic acid groups and is required for influenza virus replication.  We showed that these smaller changes can be monitored very accurately across many Uniprot and NCBI strains using hydropathicity scales to quantify the roughness of water film packages, which increases gradually due to migration, but decreases abruptly under large-scale vaccination pressures.  Here we show that, while HA evolution is much more complex, it still shows abrupt punctuation changes linked to those of NA.  HA exhibits proteinquakes, which resemble earthquakes and are related to hydropathic shifting of sialic acid binding regions.  HA proteinquakes based on sialic acid interactions are required for optimal balance between the receptor-binding and receptor-destroying activities of HA and NA for efficient virus replication.  Our comprehensive results present an historical (1945-2011) panorama of HA evolution over thousands of strains, and are consistent with many studies of HA and NA interactions based on a few mutations of a few strains.  While the common influenza virus discussed here has been rendered almost harmless by decades of vaccination programs, the sequential decoding lessons learned here are applicable to other viruses that are emerging as powerful weapons for controlling and even curing common organ cancers.  Those engineered oncolytic drugs will be discussed in future papers.

**Introduction**

A previous paper [1] showed that punctuated evolution and strain convergence occur in NA1 (H1N1) and can be identified by studying amino acid mutational changes of strains in Uniprot and NCBI databases.  The punctuations arise because of migration (interspecies, avian, swine, …, or geographical), which can increase antigenicity, and large-scale vaccination programs, which not only reverse migration effects, but go further and have led to large overall antigenic reductions.  It turned out that it was easy to monitor all these effects on NA simply by studying the global roughness of water-NA chain interfaces, using hydropathic methods previously applied to other membrane proteins, such as rhodopsin [2,3].  The success of these methods was greatly improved by the use of the MZ hydropathicity scale based on self-organized criticality (SOC) [4].  SOC explains power-law scaling, and it is arguably the most sophisticated concept in equilibrium and near-equilibrium thermodynamics.  This result shows a gratifying effective internal consistency, as SOC arises because of evolution, and its success in recognizing and quantifying punctuated evolution and strain convergence shows that long-range hydropathic interactions can dominate the more familiar short-range ionic and covalent interactions, which are the only ones treated by other methods, in optimizing evolving interactions of sufficiently large proteins.

How large must a protein be to be "sufficiently large" for long-range forces to dominate its functionality? This is a key point, especially as the goal of the present work is to develop and refine skills for engineering hybrid (mutated) Newcastle Disease Viruses (NDV, > 500 amino acids) as effective oncolytics for internal organs [5]. Methods based on short-range interactions (van-der-Waals, hydrogen bonding and continuum solvation) within Euclidean structures have emphasized chemical-protein epitope interfaces that are usually limited to 15-20 amino acids, although recently these have been expanded in yeast-based studies to 50 amino acids [6]. Because the hydropathic interactions between water films and protein substrates are weak, it has traditionally been assumed that evolutionary energy changes in short-range interaction energies always dominate long-range hydropathic energy changes. Short-range packing interactions die out for amino acid sequences longer than 9, leaving mainly water-protein roughening interactions at longer range. The packing interactions are so strong that the energies associated with them change little with tertiary conformations. The situation here is analogous to that described by the Huckel $\pi$ electron theory of the chemical activity of polycyclic hydrocarbons. The strong coplanar $\sigma$ short-range bonding interactions have little effect on most chemical properties, which are determined by the weaker $\pi$ out-of-plane bonding long-range polarization interactions [1]. The long-range dominance of hydropathic interactions is the reason that the evolution of solvent accessible surface areas follows power laws for each centered amino acid in segments of length 2N + 1 (4 < N < 17) in the survey of 5526 PDB segments, as expected from SOC and Darwinian evolution [4,7].

The 566 amino acids of compacted HA exhibit a rich Euclidean structure composed of two chains, HA 1 and HA 2 [8]. The two ends of the globularly combined chains are stabilized by hydrophobic peaks, and there is a third peak near 310 stabilizing the center of HA. See Fig. 1, which exhibits changes in the hydropathic chain profile using the MZ scale [4] averaged over a long sliding window length W = 111; this choice of W will be discussed below. The various evolutionary punctuations of companion NA sequences [1] will be examined here for HA (Fig. 1 shows one of them, the 1976 Fort Dix outbreak, accompanied by a hasty vaccination program.) The central third 310 hydrophobic peak occurs on the HA1 side of the 344 cleavage site. It stabilizes HA1 after the fusion segment 345-367 of HA2 has merged with the target membrane. As in Fig. 2 almost all the evolutionary changes occur in HA1 1-300, and this is the region discussed here in detail (see Fig. 2). It is dominated by the globular head domain, which includes the sialic acid binding site 130-230 and a number of glycosylation sites, which have increased since 1918 [8].

Uniprot lists nine glycosylation sites in HA1 (27,28,40,71,104,142,177,286, and 304) and only one on HA2 (498). Glycan microarray analysis has revealed glycosylation differences between either remote H1N1 strains (1918, modern) at a few sites [9] or differences between H1N1 and other subfamilies such as H5N1 [10]. The multiple glycosylation sites apparently contribute mainly to the HA agglutination function, as will be seen below.

**Methods**

The methods used here are almost the same as were used in [1] for NA, but with a few important changes. The evolution of NA was monitored by calculating the roughnesses of $\mathcal{R}_{KD}(W_{max})$ and $\mathcal{R}_{MZ}(W_{max})$ with $W_{max} = 17$ of the water packaging film with two scales, KD and MZ. These roughnesses (especially with the MZ scale) showed well-defined plateaus connected by punctuated decreases (vaccination programs) or increases (migration). The superior NA resolution of the MZ scale has led us to report HA results here only for the MZ scale. The sliding window width used for calculating the NA roughnesses was set at $W_{max} = 17$, although 21 (closer to the transmembrane thickness) would have been equally good. This width was also used in displaying chain hydroprofiles $<\psi(j)W>$.

The historical (1945-) directed evolution of roughnesses towards smaller values found for NA does not occur for HA. Instead, at each punctuation of NA, the HA chain hydroprofile $<\psi(j)W>$ exhibits hydrophobic stabilization over selected blocks whose edges tend to coincide with one or both of the edges of the sialic acid binding site 130-230. Because this site is so wide, we have chosen to display hydroprofiles $<\psi(j)W>$ with $W = 111$. To test this choice, we selected several HA sequences from Hawaii 2007. As was shown in [1], Hawaii 2007 contains two subsets, one labeled "Brisbane", and the other, "Solomon islands". The differences in their NA $\mathcal{R}_{MZ}(17)$ are large (30% of the historical shift from strain A to strain D*) [1], while their HA $\mathcal{R}_{MZ}(1)$ differences are only 1.5% of $\mathcal{R}_{MZ}(1)$, which is only $2\Sigma$, where $\Sigma$ is the sum of the $\sigma$'s of each subset [1]. From HA Hawaii 2007 we selected two sequences (ACB11812, Brisbane, and ACA33672, Solomon islands) with the largest BLAST non-positive differences and the largest $\mathcal{R}_{MZ}(1)$ differences. Their $<\psi(j)111>$ chain profiles exhibit a striking sign difference reversal near 180, presumed to be associated with a change in sialic acid binding (Fig. 3). A smaller window value of $W = 75$ shifts the crossover to below 180, and introduces secondary sign reversals, apparently associated with inadequate sliding window resolution (Fresnel fringes associated with sharp sialic acid edges). This shows that W = 111 is an excellent choice for HA1 sliding window width, as expected from its similarity to the length of the sialic acid binding site.

The large value of W, and the dominance of overall interactions with sialic acid on the length scale W ~ 111, explains why the details of glycosylation interactions on the length scale of glycosylation spacing (three times smaller) are important only for large-scale strain differences. A subtype-specific epitope (aa 58-72) has been found between Gly3 and Gly4 [11]. One of the sites identified as significant by glycan microarray analysis [9] is 190, which is seen in the chain hydroprofiles (Fig. 2) as one of several hydrophilic extrema.

**Results**

Perhaps the simplest difference between HA and NA profiles is the fact that NA profiles have overall $\psi_{MZ}$ averages $= <\psi_{MZ}>$ nearer 155 (hydroneutral, which is normal) than HA. This is shown in Table I, averaged over selected strains. For NA we estimated $\sigma$ from Hawaii 2011. Here for HA we find systematic parallel motion of $<\psi_{MZ}>$ and $<\psi_{KD}>$ with peaks at Fort Dix (New Jersey 1976) and swine flu (2009), as shown in Fig. 4. These correlate with the punctuations found in NA, and have a natural interpretation as the HA analogs of the NA flu pandemics, which were explained in the NA paper. Thus we can estimate the HA intrinsic scatter by evaluating $<\psi_{MZ}>$ and $<\psi_{KD}>$ in a year where there was little activity (we used 1996) and using strains widely dispersed geographically.

In Fig. 4 we can compare Table I's $\sigma$ with the Fort Dix outbreak hydrophobic peak, and the growth of the swine flu pandemic (which appeared in Brazil 2001, then grew in New York 2003, and Berlin 2005, finally reaching a peak in Texas 2007), we find that the $<\psi>$ peaks are $\alpha\sigma$ above background, with α ~ 4 for the MZ scale, and ~ 3.5 for the KD scale. This is certainly impressive, but can't we do better? Can't we connect these average changes to the chemistry of HA?

We can. We return to Fig. 1, which shows that both the A1 and A2 chains of HA are predominantly hydrophilic, with deep minima near 140. Even the relatively hydrophobic stabilizing peak near site 310 peaks near only 156. At present the reasons for this exceptionally hydrophilic HA character are not clear, but there are several possibilities. Strong interactions with water are the result of larger surface/volume ratios, and larger surfaces could facilitate cylindrical oligomer formation. This could explain why HA sequences, and not NA sequences, have been used in vaccines, where they could block oligomer formation [12,13,14].

The expanded $<\psi(j)111>$ HA1 chain profiles for the 1976 migration-vaccination punctuation are shown in Fig. 2. The large HA shift of the New Jersey (Fort Dix outbreak) sequence is expected; the Fort Dix outbreak included a large increase in the NA $\Re_{MZ}(17)$ roughness, and an even larger increase for $\Re_{KD}(17)$ [1]. More interesting is the NA shift from 1954 to 1978-1986: superstrain B (1954) reverts to superstrain A′ (1978-1986) so far as NA roughness is concerned (Table I of [1]). Here it is seen that HA progresses and becomes more flexible than NA, as the sialic acid region 120-230 becomes more hydrophilic (softer and more open) in 1978-86 than in 1954. Between 1954 and 1978 there are 10% non-positive mutations in HA1, so the effects of the Fort Dix outbreak on HA were large. There are fascinating details in the figure. For example, in 1978 we see two sharp hydrophilic minima at 130 (sialic acid site edge) and 190 (sialic acid site center), with the latter weakening by 1986. What happened from 1978 to 1986 was the KDQKTIYQK (203-211) GDQRAIYHT mutation, which includes four non-positives (KG, TA,

QH, KT), all four increasing hydrophobicity cooperatively and closing (tightening or compacting) the structure.

The first punctuation (A − B) of NA $\mathscr{R}_{MZ}(17)$ occurred in connection with the vaccination program of the American army begun in 1944, and its benefits were largest in the Netherlands 1954 sequence. For HA the punctuation itself was most apparent in the Rome 1949 – Ft. Warren 1950 sequences, as shown in Fig. 5. The largest effect was the hydrophobic plateau that occurred in the sialic acid receptor central block between 110 and 230 (sialic acid binding) in Ft. Warren 1950, but it was preceded by a smaller hydrophobic increase in the N terminal block below 130 in Rome 1949, which in turn was preceded by a still smaller hydrophilic increase in the P terminal block above 230. After the initial 1950 punctuation, the entire chain profile reverts in 1954 to nearly its 1945 state, except that there is weak hydrophilic softening of the terminal sialic acid matrix blocks below 110 and above 230. What does this pattern mean? It is mysteriously similar to an earthquake with a precursor and an aftershock, but that is not so mysterious after all. If proteins are indeed near self-organized critical states, then they might well exhibit tertiary hydropathic shocks resembling template earthquakes, as one of the first (and still perhaps the most popular, 640 papers) applications of SOC has been to earthquakes [14-16], where collisions of tectonic plates are described by spring-block models [14]. These quakes occur in the water film packaging HA1, and resemble earthquakes in the earth's crust. Growing actin networks also exhibit sporadic effects predicted by SOC [17]. Mechanical spring-block models of magnetization phenomena and Barkhausen noise also reproduce microscopic pictures of domain wall networks' movement and pinning [18]. These structural changes can be described as proteinquakes.

The 1989-2003 NA smoothing gain ended with the advent of the "swine flu" strains, which appeared first in Hong Kong and next in New York in 2003 and in Berlin in 2005. The characteristic HA feature of these strains was a large hydrophobic increase in the N terminal sub-130 block, which was almost identical for New York and Berlin (Fig. 6). These hydrophobic block increases correspond to block elastic stiffening. As shown in Fig. 7, comparison with an actual 2007 swine flu sequence shows that the latter sequences were also evolving rapidly. How did swine flu evolution compare with human flu evolution?

As shown in Fig. 8, a new superstrain of swine flu first appeared in Hong Kong in 1999. The England 1998 strain was very different, being much more hydrophobic in the post-230 block, and much more hydrophilic in the silaic acid block 130-230 (presumably less antigenic). Hong Kong 1999 swine flu quickly spread to North Carolina 2000, but by 2003 its increased 130-230 antigenicity had been halved, and the increase in the sub-130 block had nearly disappeared. Thus the new flu strain appeared in swine several years before its first human appearance in New York in 2003, and had been controlled in swine before the new strain had become a human problem, which probably limited its human impact.

With a larger data base, we can study the response to the 2007 swine flu vaccination in one locality, which was done for North Carolina and for Norway. The results for North Carolina, see Fig. 9, show a large shift towards hydrophilicity between 2008 and 2009. The results for Norway are similar, but with an interesting difference. In 2006 there were 4 non-positive BLAST mutational differences between North Carolina and Norway, but by 2009 the two sequences had converged to become identical. This convergence probably occurred as a result of viral evasion of a common vaccination program, which was so effective as to erase substantial viral climatic differences. A simple search of the "identical proteins" feature of the HA NCBI data base, similar to the previous one for NA [1], is less rewarding for HA because of a wider range of reported HA lengths. However, one length (566 aa) is sufficiently common in the 2009 and 2010 HA data to be indicative of effects similar to those found for NA: a single very common strain in 2009, which becomes less common in 2010 as vaccination pressures receded. According to [1], the effects of the swine flu vaccination program shown there on human NA are remarkable, as the human strains made a very large "Lévy" jump from superstrain C to superstrain D (not seen in earlier punctuations) to avoid the vaccine and dodge swine flu, while reducing their severity. Apparently the large hydrophilic softening seen in Fig. 8 is a Lévy effect for HA. These two large Lévy effects in sequence space, which describe explicitly smoothing of NA and block softening of HA, are quantitative measures of the optimal balance between receptor-binding and receptor-destroying activities of NA and HA that is required for efficient virus replication [19].

These large jumps in HA block hydropathicity are mediated by strings of mutations, which were not seen in NA. Also NA contracted slightly and stiffened in response to the swine flu vaccination program, while HA expands and softens – another example of NA-HA "balance" [19]. One is not surprised to find that HA has a block structure, because it binds to the cell that is being infected through silidase, which is a large molecule. Also it seems natural that the large number of HA mutations induced by the swine flu vaccination program should occur in short strings (see Figure captions), rather than as largely isolated single mutations (as in NA), as this is a more effective way to alter block hydropathic structure, with fewer changes in short-range packing [18].

**Conclusions**

The analysis of NA using hydropathic roughness with W ~ 20 as a configuration coordinate revealed opposing punctuations due to migration (increasing roughness) and vaccination programs (increasing smoothness). Here we have found that balanced interactions between HA and NA also cause parallel punctuations in HA at the same times. However, the HA punctuations are more complex, and require studying hydropathic chain profiles with large sliding window lengths of order the sialic acid binding site size, W ~ 100. The HA1 profiles have a characteristic broad and deep (butterfly) hydrophilic minimum in the sialic acid binding region 130-230, and are strongly hydrophilic. The overall evolution of HA1 shows primarily shifts in hydropathicity of three chain blocks, 50-130, 130-230, and 230-300. These punctuated block shifts closely resemble earthquakes [14-16] in one dimension, consistent with analyzing them using the MZ hydropathicity scale based on self-organized criticality. Unlike NA roughness shifts, which are often associated with a few single amino acid mutations, HA block shifts are often associated with sequences of up to 9 amino acids, which reflect the epitopic chemistry [13] of short-range HA-sialidase interactions. The hydropathic chain profiles can be used as biomarkers to represent HA-sialidase interactions.

The central limitation of prior studies of viral kinetics has been their low resolution, limited to the large N-glycan spacing length scale [9,10] or the even larger sialic acid length scale [22]. There one finds evidence that N-glycans guide partner ligands to their binding sites and prevent irregular protein aggregation by covering oligomerization sites away from the ligand-binding site [23]. Here we have shown that detailed hydropathic chain profile analysis enables resolution of punctuated evolution at the level of individual glycoprotein amino acids (about 20 times smaller than N-glycan spacing). Why are proteinquakes so important at the molecular level? Formation of oncolytic core oligomers requires multiple steps. First, individual viruses must be bound to the cancer cell membrane. Next these isolated molecules must diffuse along the membrane surface to form oligomeric clusters through shear flow [24]. The surface diffusion rate over a rough surface can be accelerated by smoothing both the NA and HA glycoproteins, another reason for viral glycosidic balance [25]. Combining HA and NA into oligomeric complexes itself involves multiple conformation changes that are expected to be dominated by π-like sliding conformational dynamics, not compressive σ-like sliding. Finally, the HA analysis leads to a block model of the HA water packaging structure that includes HA fractures (Fig. 4) which could be called vaccination-driven proteinquakes. Such cumulative mechanical effects are even the cause of stiff α helix/soft β strand transitions in 56 aa proteins that share 88% sequence identity [26]. Here in 577 aa proteins the sequence identities are even higher, Rome 1949-Ft Warren 1950: 94% identity (ABN59434-Q288V2). Mechanical effects dependent on stiff/soft alternation are a common theme in structures like bone, which are strong but not brittle (high yield strength). They are one of the dominant factors in ligand binding [27].

The success of this historical (1945-2011) panorama depends almost entirely on the large data base that has accumulated. The data have come from many sources, but the largest part of this data base, especially the older parts, is due to the NIAID Influenza Genome Sequencing Project, which has proved to be invaluable. The present analysis has important implications for engineering oncolytically superior strains of Newcastle disease virus for cancer therapy [28], which will be presented separately.

|            | <MZ>  | σ(<MZ>) | <KD>  | σ(<KD>) |
|------------|-------|---------|-------|---------|
| Lyso (C)   | 153.1 |         |       |         |
| Lyso (H)   | 154.7 |         |       |         |
| Adren (H)  | 154.7 |         |       |         |
| Rhodop (L) | 167.1 |         |       |         |
| Rhodop (C) | 167.6 |         |       |         |
| Rhodop (H) | 167.8 |         |       |         |
| NA         | 151.5 | 0.4     | 159.7 | 0.3     |
| HA         | 148.9 | 0.6     | 156.5 | 0.7     |

Table I. Strain-averaged panoramic average hydropathicities $<\psi>$ for the two scales in the text. The standard deviations σ for NA are taken over the panorama plateaus 1945-2011. For HA the year 1996 was used to estimate σ (see text). Also shown are lysozyme *c*, adrenergic (β1) and rhodopsin values for several species (Lamprey, Chicken, Human). In general evolution stabilizes proteins by compacting them and increasing $<\psi>$. Note

that rhodopsin is exceptionally stable, as it must be to receive and process optical signals. NA is noticeably hydrophilic, and HA is even more hydrophilic. The $<\psi>$ trends shown here are interesting, but the opposing viral effects of migration and vaccination pressures can be recognized only in the context of the more sophisticated discussion in the text.

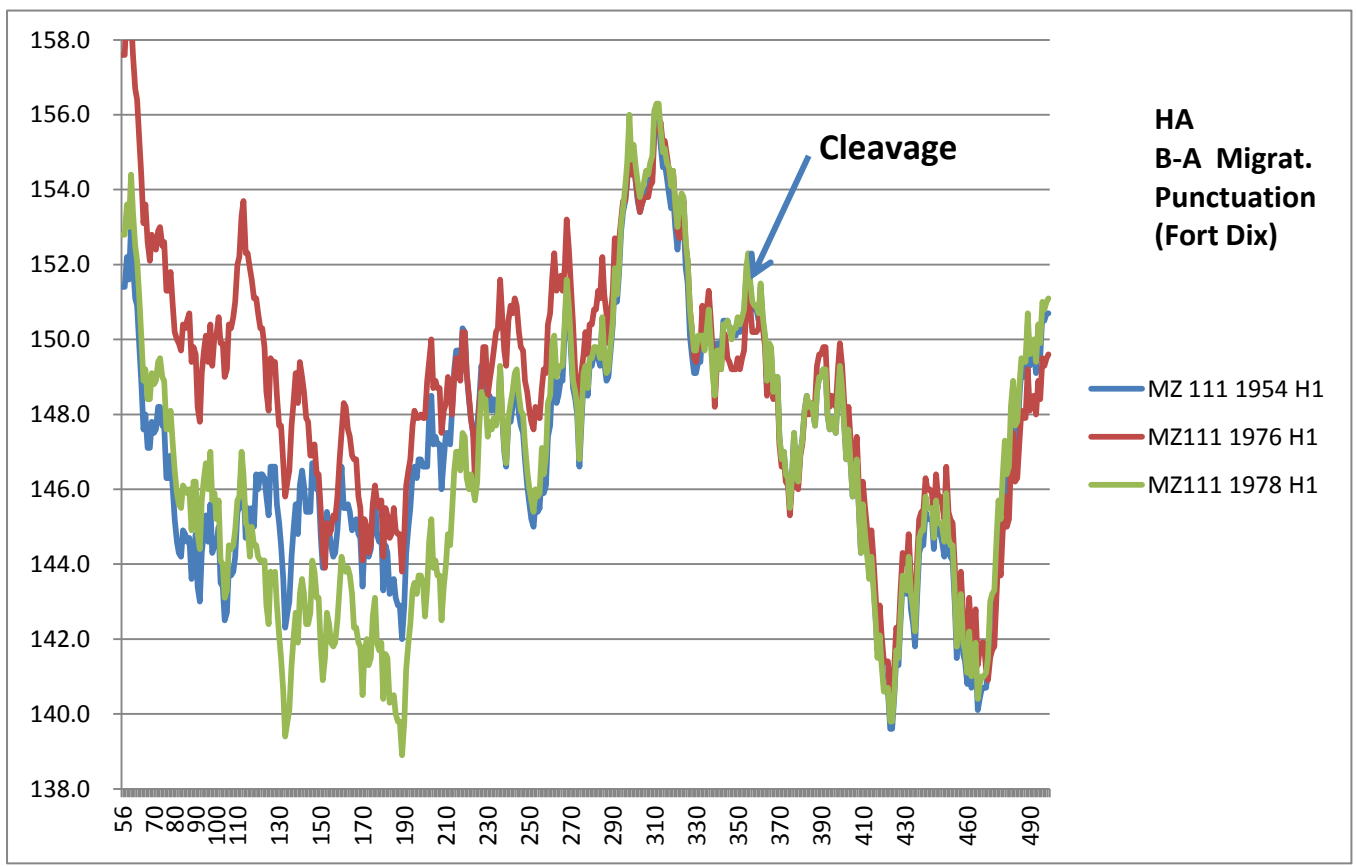

Fig. 1. The full $<\psi(j)111>$ chain profiles of three HA sequences, Netherlands 1954, Fort Dix outbreak (1976), and post-vaccination California 1978. The cleavage site separates HA1 from HA2. Mutations occur primarily in the HA1 region below 300.

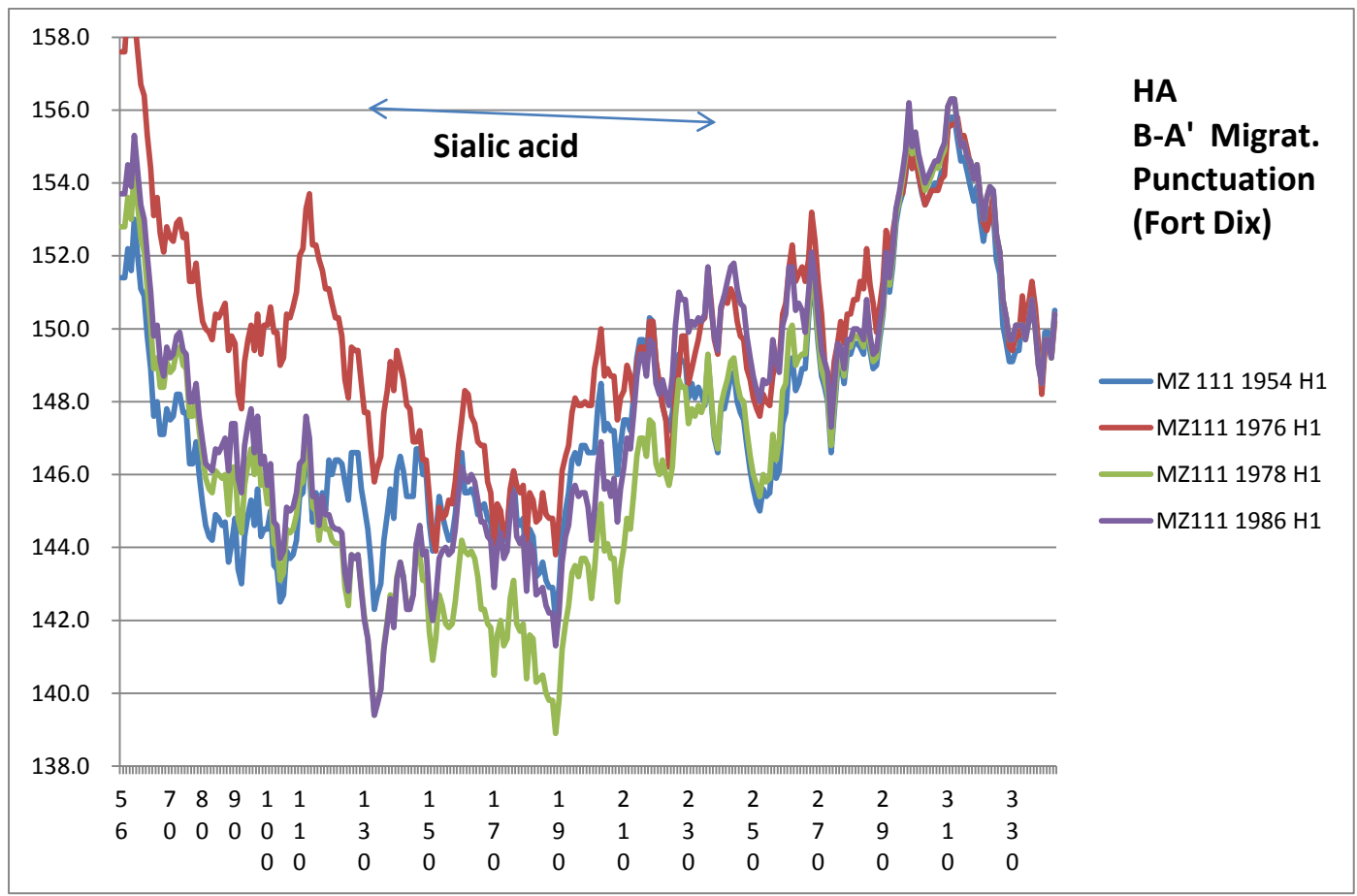

Fig. 2. The <ψ(j)111> chain profiles of four HA sequences, Netherlands 1954 (ADT78876), Fort Dix outbreak 1976 (ACU80014), post-vaccination California 1978 (ABY81349, identical to Memphis 1983), and Texas 1986 (ABO44123), cut off at 330 (A1 chain only) to show more clearly mutational trends below 300. The Fort Dix outbreak increased hydrophobicity, especially below 150. After the 1976-77 vaccination program, in 1978-86 hydrophobicity dropped below 1954 in the region 120-230, essentially the sialic acid binding site.

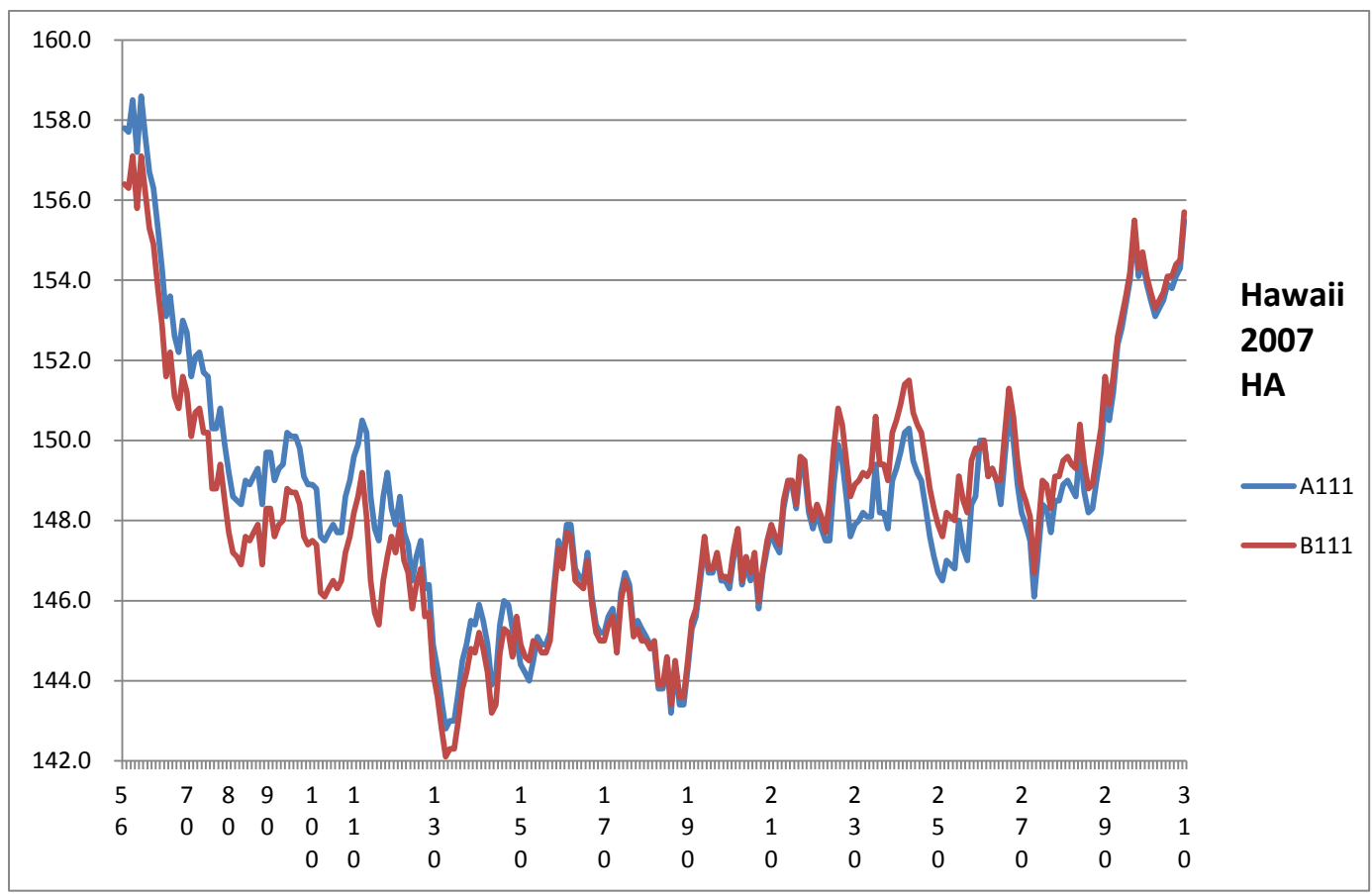

Fig. 3. The <ψ(j)111> chain profiles (55 < j < 311) of Genbank Brisbane ACB11812 (B) and Solomon islands ACA33672 (A) [Hawaii 2007]. The key B-A mutations are K64I, MT(205,206)KA and N238D. The crossover near 180 can be described as a hydropathic twist.

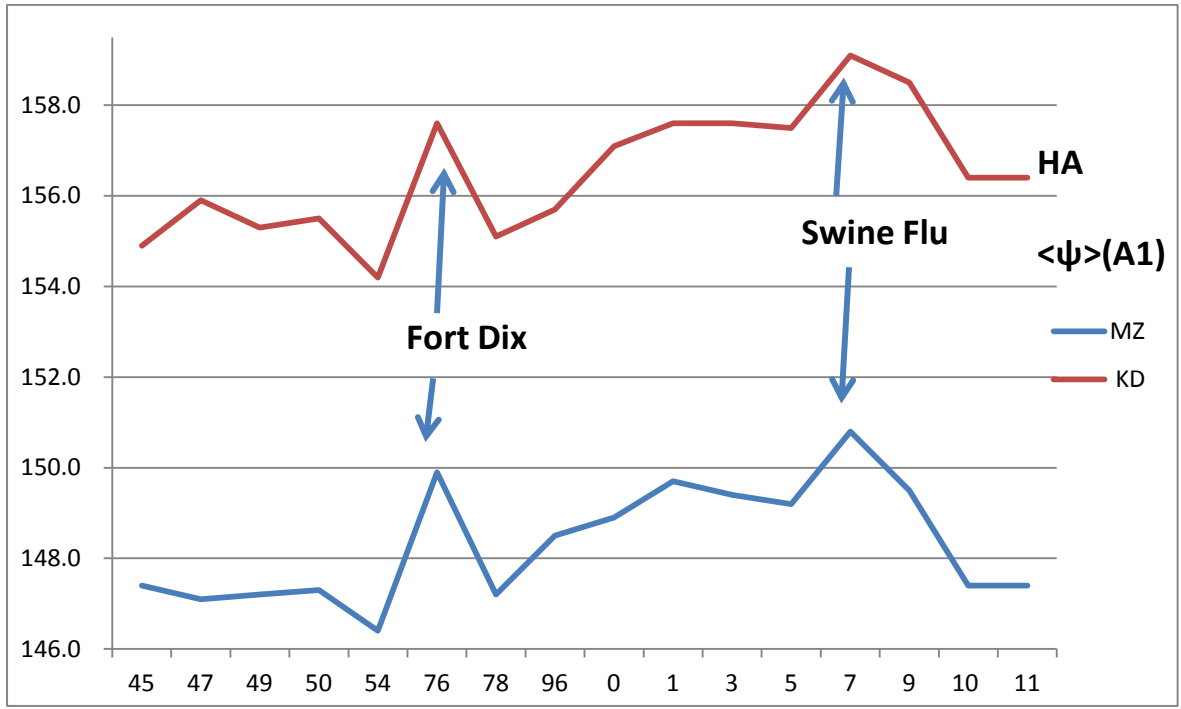

Fig. 4. Sketch of chain A1 <ψ(aa,1)> evolution, with Fort Dix outbreak and swine flu peak (before vaccination program) indicated. The HA <ψ(aa,1)> trends seen here parallel those for NA tabulated in [5].

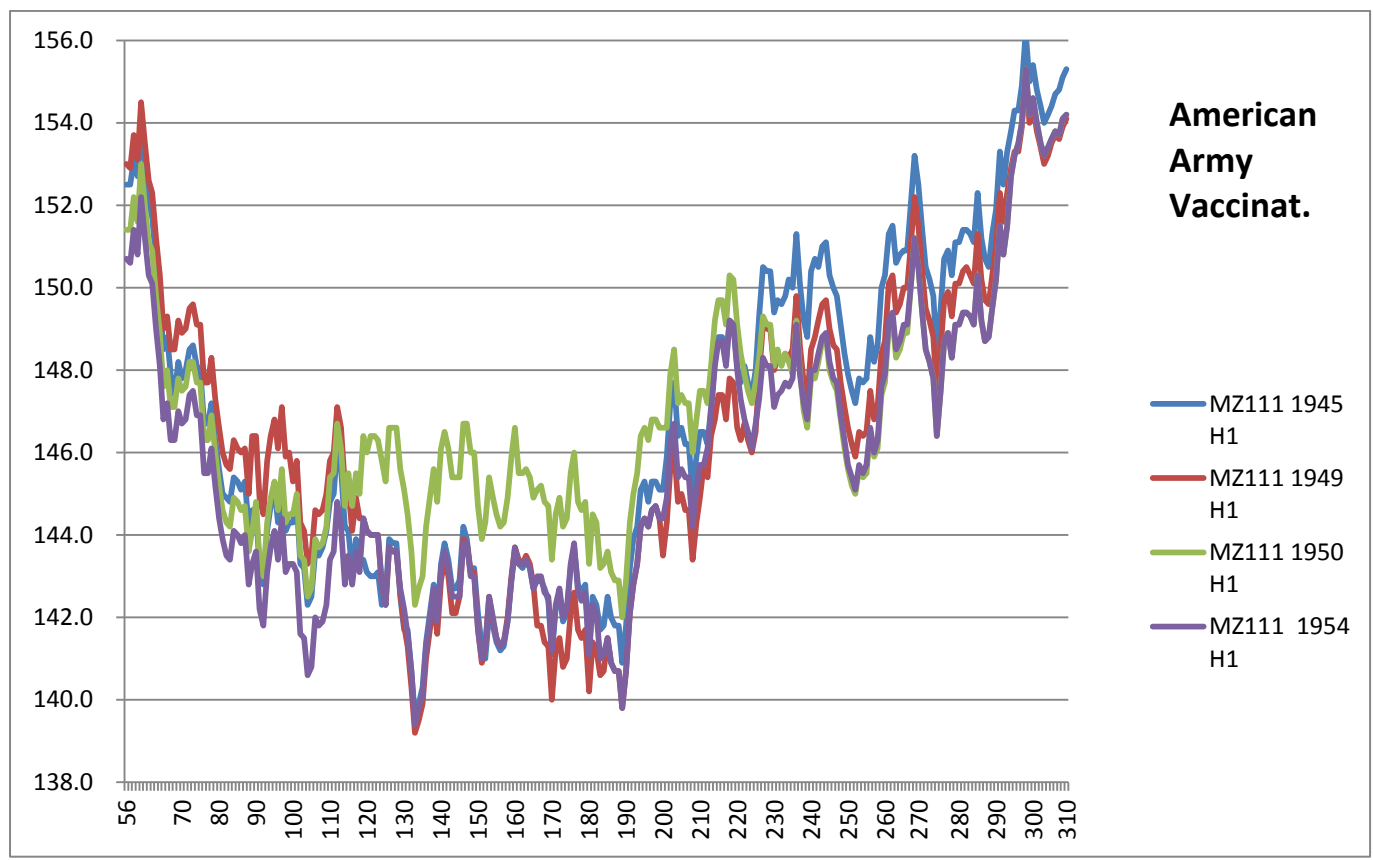

Fig. 5. The Army vaccination punctuation chain profiles resemble an earthquake with respect to the sialic acid binding site 130-230 relative to its HA1 matrix (see text). Note that the main feature of the sialic acid binding site is not the hydrophilic extrema at its end points, but rather the flattening of the HA butterfly profile across the entire 130-230 binding range. This reflects the strongly one-dimensional nature of water film packages, which is not obvious in Euclidean simulations of protein dynamics.

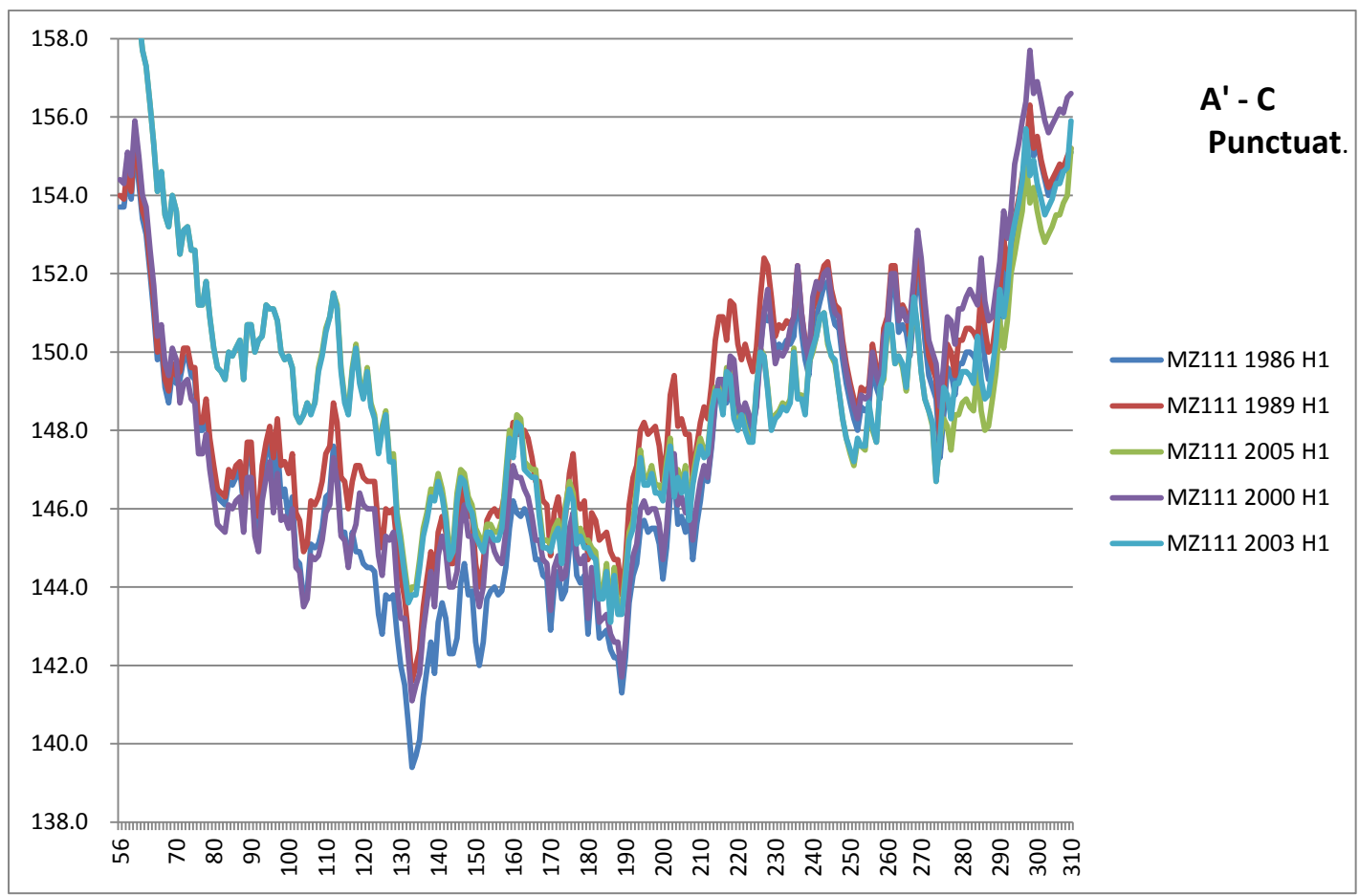

Fig. 6. During the period 1983-2005 NA exhibited a plateau that terminated with punctuation B-C (advent of swine flu), which occurred in Brazil (AAY42117) in 2001, New York in 2003 and in Berlin in 2005. The advent of swine flu caused the HA1 block below 130 to increase hydrophobically almost identically in New York in 2003 and in Berlin in 2005, as shown here. The Genbank sequences used here are ABO44123 (Texas 1986), ACL 12261 (Siena 1989), ACI32714 (Berlin 2005), AAX56530 (New York 2000) and ABB82205 (New York 2003).

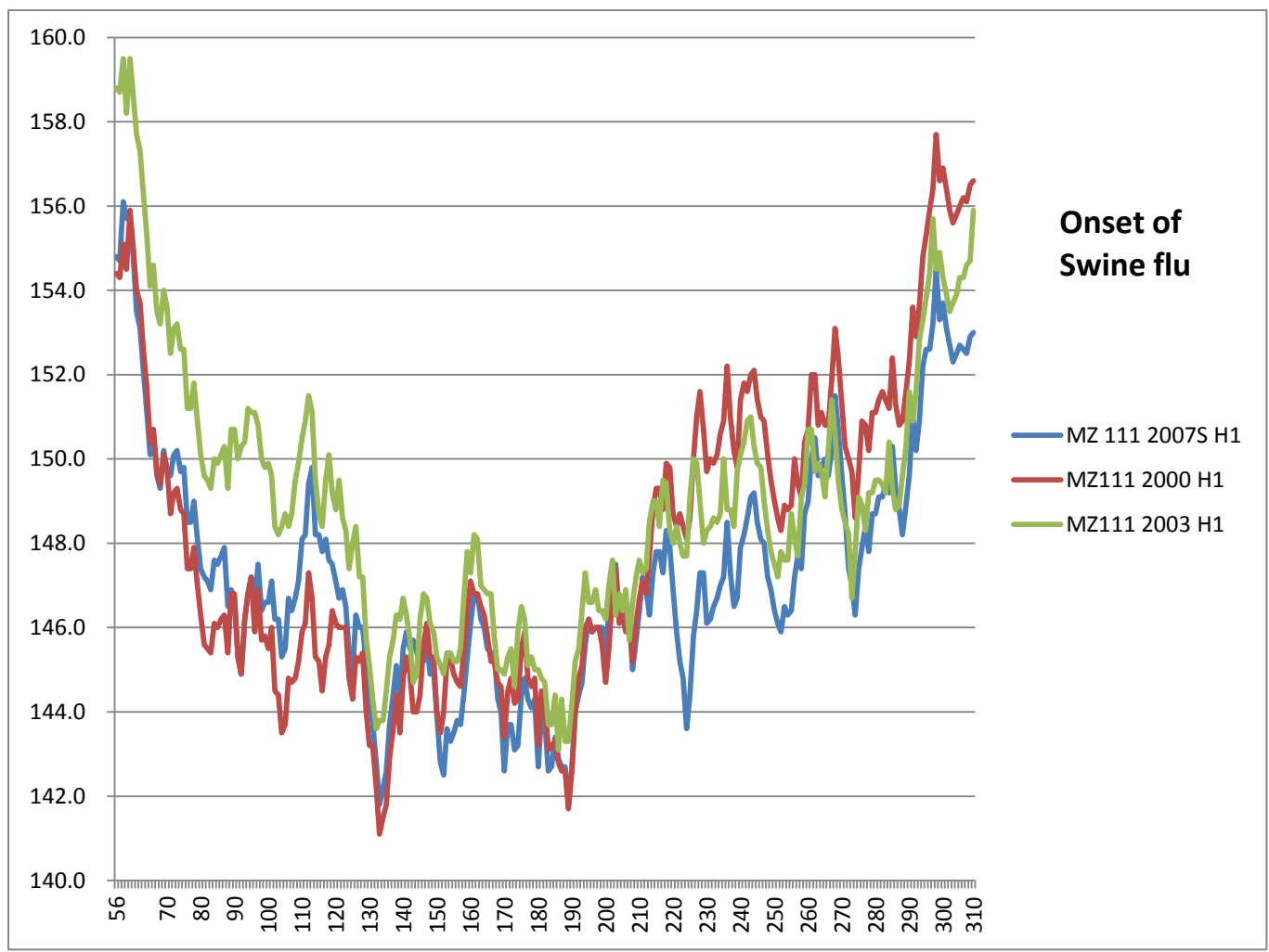

Fig. 7. Comparison of human strains from New York 2000 and 2003 with swine flu from Kansas 2007. The similarities are obvious, but what actually happened is described in Figs. 8 and 9.

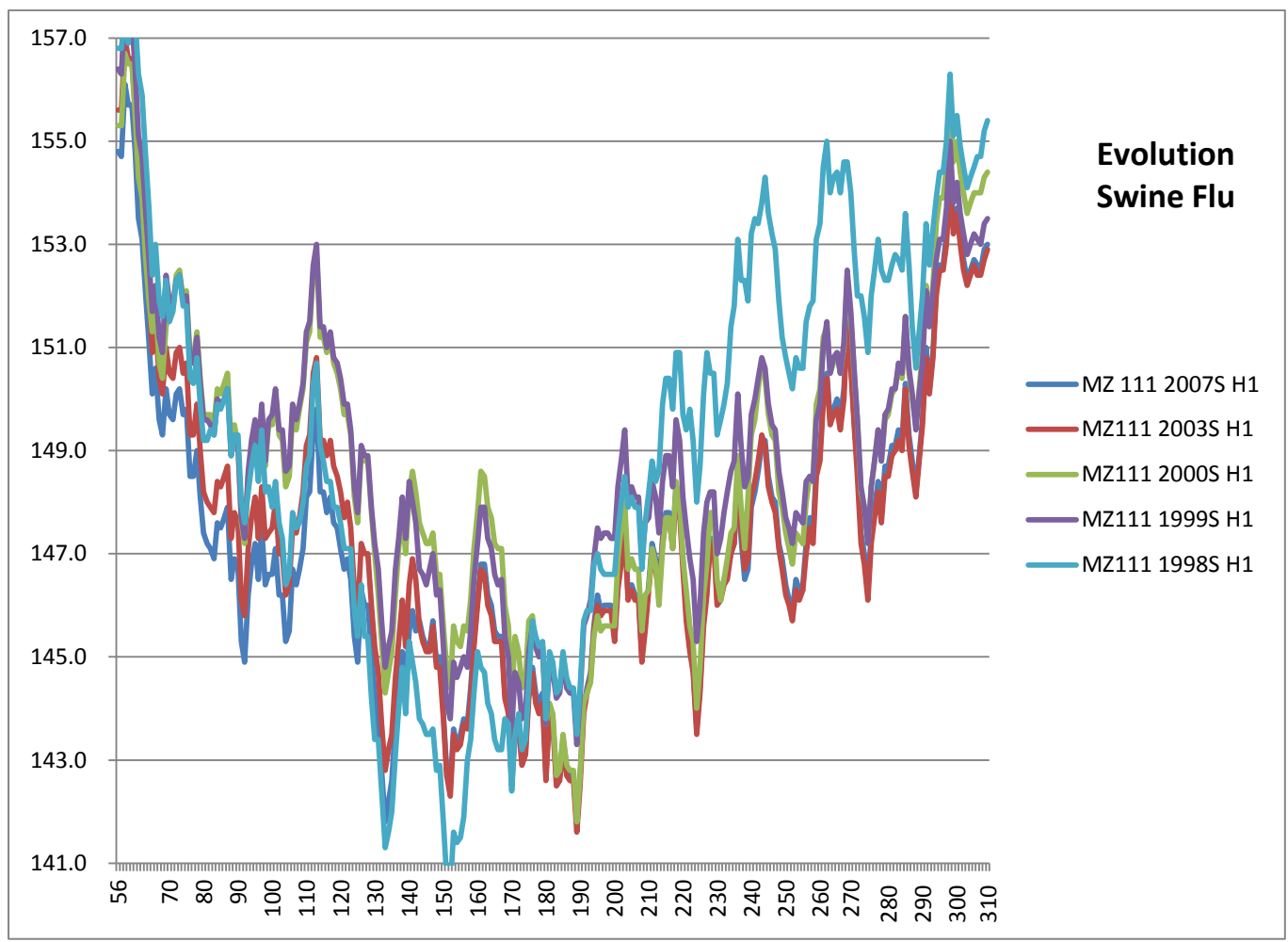

Fig. 8. The swine flu sequences profiled here are England 1998, Hong Kong 1999, North Carolina 2000 and 2003, and Kansas 2007.

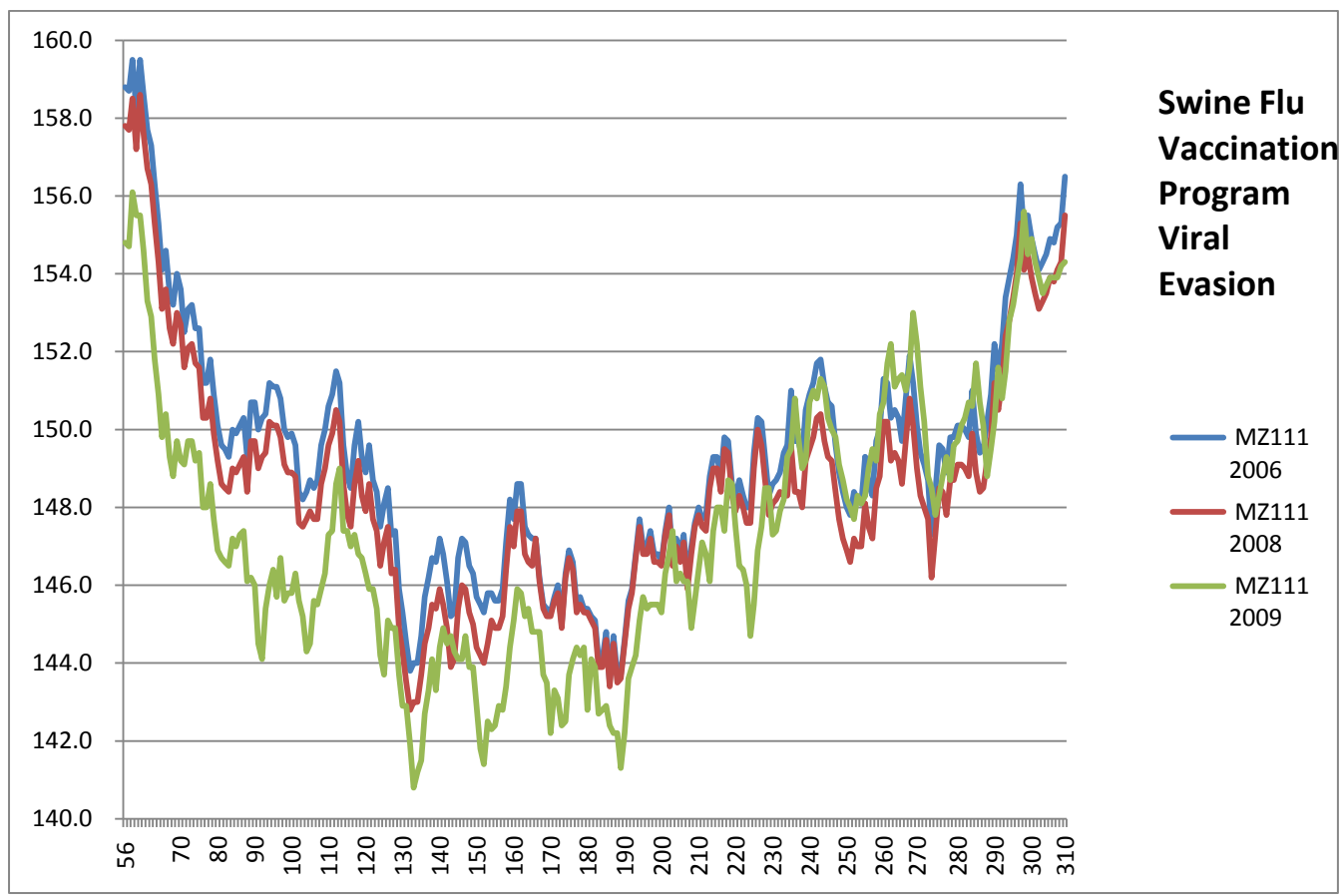

Fig. 9. North Carolina response to swine flu vaccination program, which began in 2007. The small response in 2008 is greatly enhanced by 2009, after which changes were minimal. The key mutations from 2008 (ACD45795) to 2009 (ADM21399) often involve not individual sites, but short strings, for instance, TATY13-16ATAN, LLISKE86-91SLSTAS, and TVT144-147DSNK, indicative of both long-range hydrophilic softening and short-range expansion, including the hydrophilic insertion 144D.